\documentclass[aps,pra,twocolumn,preprintnumbers,amsmath,amssymb,floatfix,showpacs]{revtex4-1}%

\usepackage{graphics}
\usepackage[pdftex]{graphicx}
\usepackage{subfigure}
\usepackage{rotating}
\usepackage{mathrsfs}
\usepackage{amsfonts}
\usepackage{amsmath}
\usepackage{amssymb}
\usepackage{placeins}
\usepackage{bm}

\usepackage{soul}
\usepackage{fixmath}
\usepackage{physics}

\usepackage[colorlinks]{hyperref}
\usepackage{siunitx}

\newcommand{\transp}{\ensuremath{\mathrm{T}}}

\DeclareMathOperator{\diag}{diag}

\begin{document}

\title{Hidden and detectable multimode squeezing from micro-resonators}
\author{ \'{E}lie Gouzien$^{1,2}$, Laurent Labont\'{e}$^{2}$, Alessandro Zavatta$^{3,4}$, Jean Etesse$^{2}$, S\'{e}bastien Tanzilli$^{2}$,  Virginia D'Auria$^{2,5}$, Giuseppe Patera$^{6}$}
\affiliation{
$^1$ Universit\'{e} Paris--Saclay, CNRS, CEA, Institut de physique th\'{e}orique, 91191, Gif-sur-Yvette, France\\
$^2$Universit\'{e} C\^{o}te d'Azur, CNRS, Institut de Physique de Nice, Parc Valrose, 06108 Nice Cedex 2, France\\
$^3$ Istituto Nazionale di Ottica (INO-CNR) Largo Enrico Fermi 6, 50125 Firenze, Italy\\
$^4$ LENS and Department of Physics, Universit\`a di Firenze, 50019 Sesto Fiorentino, Firenze, Italy\\
$^5$ Institut Universitaire de France (IUF), France\\
$^6$Univ. Lille, CNRS, UMR 8523 - PhLAM - Physique des Lasers Atomes et Mol\'{e}cules, F-59000 Lille, France}


\begin{abstract}
In the context of quantum integrated photonics, this work investigates the quantum properties of multimode light generated by silicon and silicon nitride micro-resonators pumped in pulsed regime. 
The developed theoretical model, performed in terms of the morphing supermodes, provides a comprehensive description of the generated quantum states. 
Remarkably, it shows that a full measurement of states carrying optimal squeezing levels is not accessible to standard homodyne detection, thus leaving hidden part of generated quantum features. 
By presenting and discussing this behaviour, as well as possible strategies to amend it, this work proves itself essential to future quantum applications 
exploiting micro-resonators as sources of multimode states.
\end{abstract}


\maketitle

Silicon (Si) and Silicon Nitride (SiN) quantum photonics offer a precious possibility to propel practical quantum optical technologies thanks to 
high density integration of high-performance functions over small footprint chips~\cite{Wang2020}. 
In recent years, a particular interest has been driven by the possibility of exploiting their optical nonlinearities to generate on-chip highly multimode entanglement among frequency-time modes. 
Four-wave mixing (FWM) in silicon-based rings or disk-shaped micro-resonators have been used to demonstrate chip-scale sources of paired-photons~\cite{Oser,Imany,Kues2022}, 
low-dimension quantum frequency combs~\cite{Morandotti2019} and, more recently, two-colour intensity-~\cite{Dutt2015} and quadrature-~\cite{Vaidya2019,Gaeta2020,PfisterCombSi2021} 
entanglement in continuous variable (CV) regime. 

Most of realisations and reported theoretical models refer to Si and SiN resonators pumped in continuous wave regime~\cite{Chembo2016,Guidry2022}. 
This theoretical paper rather focuses on the study of multipartite states produced by micro-resonators pumped by optical pulses, as a successive natural step toward more complex architectures. 
Beside lower oscillation threshold, this regime offers multimode entangled states that exhibit a way richer structure~\cite{Fabre2020} 
as well as the possibility of tailoring their features~\cite{Patera2012,Arzani2018}. This works focuses in particular on CV frequency-time entanglement~\cite{Valcarcel2006,Patera2010}, 
due to its important applications in quantum metrology~\cite{Fabre2020}, quantum communication~\cite{Dellantonio2018} 
and measurement-based quantum computing~\cite{Menicucci2007}. 
The presented characterisation of the non-classical properties of micro-resonators is performed in terms of morphing supermodes, 
mapping the full dynamics of multipartite states into that of independent single-mode squeezed states whose spectral shape depends on a continuous parameter~\cite{Gouzien2020}.
Such an analysis reveals that in standard  working conditions, a full characterisation of CV quantum properties out of micro-resonators is not accessible 
to traditional quadrature homodyne detection, thus leaving optimal squeezing features hidden. 
This aspect is analogous to what observed for quantum states whose noise spectra are asymmetrical with respect to the carrier~\cite{Barbosa2013a,Barbosa2013b} and produced, \emph{e.g.}, by resonant phenomena such as atomic emission. Nevertheless, it has never been high-lightened by former works on silicon-based micro-resonators. By showing and discussing it, this work anticipates difficulties that may occur in experiments involving micro-resonators as sources of multipartite squeezing. In addition, it identifies possible system engineering strategies leading to configurations where the problem is less severe. Its impact is, thus, essential for the conception and future experimental realisations of quantum technologies applications 
exploiting pulsed multimode states.
\\

\begin{figure}[t!]
\includegraphics[width=0.9\columnwidth]{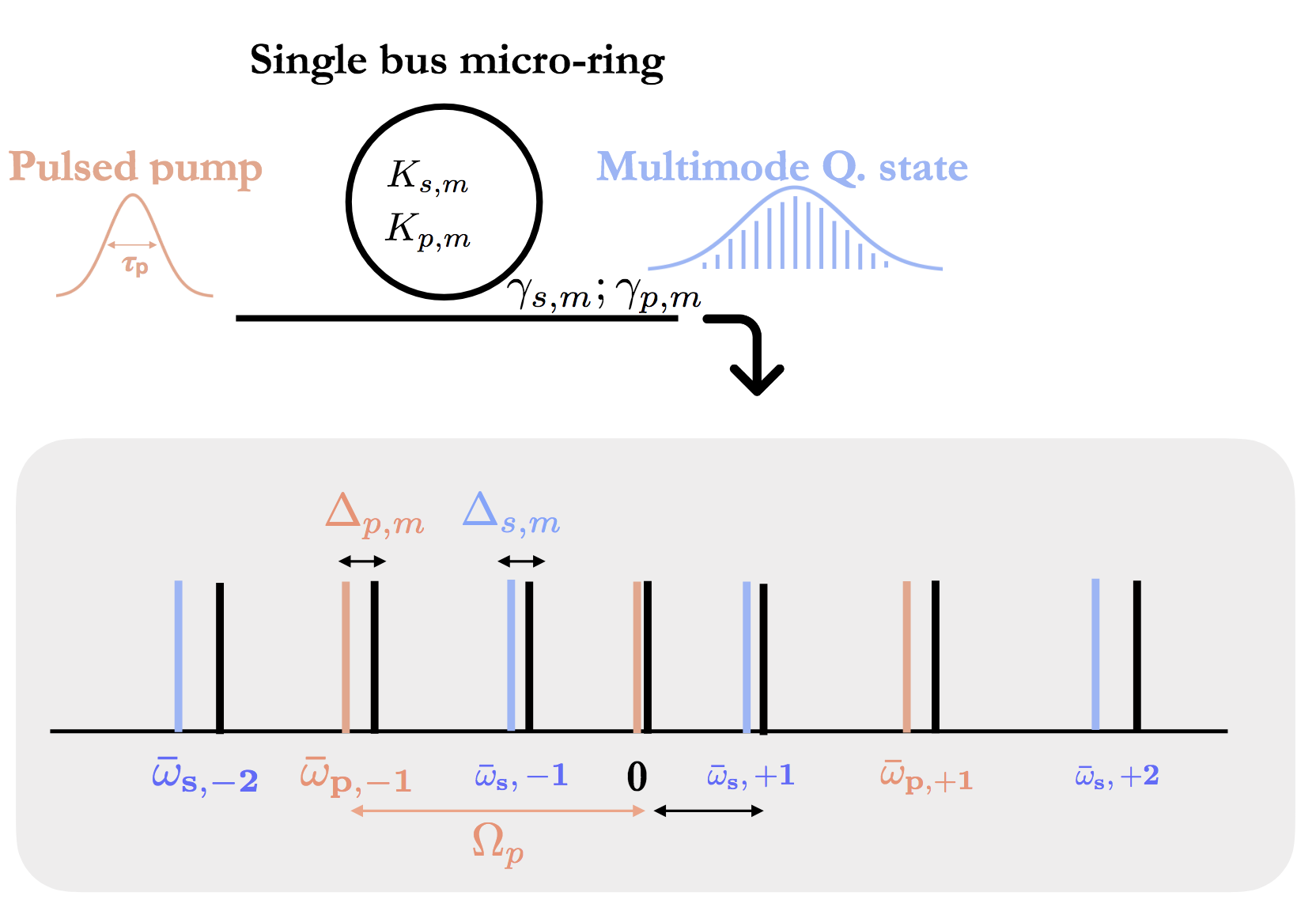}
\caption{Schematic of the considered nonlinear interaction. The pump is taken to be a frequency comb with carrier frequency $\omega_{\mathrm{p}}$, pulse duration $\tau_{\mathrm{p}}$ and 
repetition rate $\Omega_{\mathrm{p}}$ approximately equal to the double of the cavity FSR. This means that the spectral components of the pump match one cavity resonance out of two, 
each with a possible detuning $\Delta$. Frequency entangled modes are emitted in the micro-resonator resonances free from the pump spectral components.}\label{Figure1}
\end{figure}

\textbf{Synchronously pumped microrings}.
As shown in  Fig.~\ref{Figure1}, without loosing in generality, the system here investigated is a micro-resonator coupled to a single straight injection waveguide (single-bus device) 
and pumped by an infinite train of optical pulses. Pump pulses are taken to have a duration $\tau_{\mathrm{p}}$ and a
repetition rate $\Omega_{\mathrm{p}}$, corresponding, in frequency domain, to a comb of equally spaced spectral components $\bar{\omega}_{\mathrm{p},m}=\omega_{\mathrm{p}}+m\Omega_{\mathrm{p}}$ 
($m$ being an integer), spanning over a range $\sigma_{\mathrm{p}}\propto 2\pi/\tau_{\mathrm{p}}$ around the optical carrier at frequency $\omega_\mathrm{p}$. 
In order to address a particularly common experimental situation, a type-0 FWM process is considered for squeezing generation. This choice of phase-matching condition gives access 
to high nonlinear conversion efficiency and it is thus particularly well compatible with the CV regime~\cite{Vaidya2019,Gaeta2020,PfisterCombSi2021}. 

In the frequency domain, the FWM interaction modes are determined by frequencies corresponding to the cavity resonances~\cite{Chembo2016}:
\begin{equation}
\omega_{m}=
\omega_0+\sum_{k\geq 1}\frac{\Omega_k}{k!}m^k.
\label{resonanceTaylor}
\end{equation}
The reference label $m=0$ indicates the resonance whose frequency approximately matches the pump carrier, $\omega_0\approx\omega_\mathrm{p}$ (see Fig.~\ref{Figure1}).
The first order parameter $\Omega_1=c/(n_g R_{\mathrm{eff}})$ gives the average cavity free spectral range (FSR) as a function of the speed of light in vacuum
$c$, the group index $n_g$, and of the ring effective radius $R_{\mathrm{eff}}$~\cite{VienBook}. In order to have distinct signal and pump modes and guarantee a synchronous pumping regime, 
the pump repetition rate is taken to be $\Omega_\mathrm{p}\approx2\Omega_1$ (\emph{i.e.}\@ $\bar{\omega}_{\mathrm{p},m}\approx\omega_{\mathrm{p}}+2m\Omega_1$). 
Note that, in general, the pump injection components approximately match one cavity resonance out of two. 
Their detuning with respect to cavity resonances $\Delta_{\mathrm{p},m}=\omega_{2m}-\bar{\omega}_{\mathrm{p},m}$ changes with $m$ due to dispersion. 
Its effect is included in the parameter $\Omega_{2}=-(n'_g c^ 2)/(n_g^ 3 R_{\mathrm{eff}}^ 2)$ that accounts for second-order dispersion effects via the frequency derivative $n'_g$ 
together with higher-order dispersion terms $\Omega_{k>2}$. As depicted in the Fig.~\ref{Figure1}, frequency entangled signal modes are generated by FWM at frequencies 
$\bar{\omega}_{\mathrm{s},m}=\omega_{\mathrm{p}}+(2m+1)\frac{\Omega_{\mathrm{p}}}{2}$ (``s'' stands for ``signal'') and can thus be unequivocally distinguished from the pump. 
They are in general detuned by $\Delta_{\mathrm{s},m}$ with respect to the odd cavity resonances.\bigskip

\textbf{Multimode linear quantum Langevin equations.}
Quantum properties of multimode light out of the micro-resonator are obtained by solving a system of coupled Langevin equations~\cite{Valcarcel2006,Patera2010}. 
These describe the evolution of bosonic operators associated to the pump ($\hat{p}_m$) and signal ($\hat{s}_m$) intra-cavity modes (``m'' being the mode label) 
and can be derived from the system Hamiltonian by following the prescriptions for fields quantization in dispersive dielectric materials~\cite{Drummond1999,Quesada2017,Raymer2020}. 
All the details on the derivation of Langevin equations and of their elements are reported in the Appendix. 

Langevin equations are linearized around a stable classical stationary solution where the pump modes are macroscopically populated, 
$\langle\hat{p}_m\rangle\neq0$, and the signal modes are empty, $\langle\hat{s}_m\rangle=0$. This corresponds to micro-resonators below their oscillation threshold. 
Note that, the $\langle\hat{p}_m\rangle$ depend on the injected pump power $P$, on the  detuning $\Delta_{p,0}$, and on the FSR-mismatch $\Delta\Omega=\Omega_1-\Omega_{\mathrm{p}}/2$. 
Linear Langevin equations can be conveniently expressed in terms of the amplitude and phase quadratures of the signal modes, 
$\hat{x}_m=(1/\sqrt{2})(\hat{s}_m^\dag+\hat{s}_m)$  $\hat{y}_m=(\mathrm{i}/\sqrt{2})(\hat{s}_m^\dag-\hat{s}_m)$. In a compact matricial form: 
\begin{align}
\frac{\mathrm{d}\hat{\mathbf{R}}(t)}{\mathrm{d}t}&=
(-\Gamma+\mathcal{M})\hat{\mathbf{R}}(t)+
\sqrt{2\Gamma}\,\hat{\mathbf{R}}_{\mathrm{in}}(t),
\label{langevin quad}
\end{align}
where 
$\hat{\mathbf{R}}(t)=(\hat{\mathbf{x}}(t)|\hat{\mathbf{y}}(t))^\transp$ is the column vector of the intracavity mode quadratures 
$\hat{\mathbf{x}}(t)=(\ldots,\hat{x}_{-1},\hat{x}_{0},\hat{x}_{+1},\ldots)^\transp$ and $\hat{\mathbf{y}}(t)=(\ldots,\hat{y}_{-1},\hat{y}_{0},\hat{y}_{+1},\ldots)^\transp$ 
while $\hat{\mathbf{R}}_{\mathrm{in}}(t)$ the quadratures of the input signal modes, here set in the vacuum states to describe a spontaneous interaction. 
The diagonal matrix $\Gamma$ describes mode-dependent coupling losses of the single-bus cavity.
Propagation losses can be included in $\Gamma$ (see Appendix and ~\cite{TheseElie}).
The interaction matrix $\mathcal{M}$ is expressed as
\begin{equation}
\mathcal{M}=
\left(
\begin{array}{c|c}
\mathrm{Im}\left[G+F\right] & \mathrm{Re}\left[G-F\right]
\\
\hline
-\mathrm{Re}\left[G+F\right] & -\mathrm{Im}\left[G+F\right]^\transp
\end{array}
\right)
\label{eMMe},
\end{equation}
in terms of the complex matrices $G$ and $F$ (with $G=G^\dagger$ and $F=F^\transp$\footnote{We are using the following notation: ${[\cdot]}^\transp$ for the transpose, ${[\cdot]}^*$ 
for the complex conjugate and ${[\cdot]}^\dagger$ for the Hermitian transpose.}). 
Matrix $G$ contains mode-dependent detunings and all terms accounting for self- and cross-phase modulation (referred here as nonlinear dispersion terms), 
while $F$ accounts for parametric amplification processes. For the micro-resonator systems considered in this work, their elements are:
\begin{align}
F_{m,n}=&{g}
\sum_l
\langle\hat{p}_{m-l+n+1}\rangle\langle\hat{p}_l\rangle,
\label{F}
\\
G_{m,n}=&
\Delta_{\mathrm{s},m}\delta_{[m-n]}
+
g
\sum_l
2\langle\hat{p}_{m+l-n}\rangle\langle\hat{p}_l\rangle^*.
\label{G}
\end{align}
The elements of both matrices explicitly depend on the pump stable steady states.
In the previous expressions, $\delta_{[m-n]}$ is the Kronecker delta and $g$ is the nonlinear strength. 
The quadratures $\hat{\mathbf{R}}_{\mathrm{out}}$ of modes at the micro-resonator output can be obtained via input-output relations 
$\hat{\mathbf{R}}_{\mathrm{in}}+\hat{\mathbf{R}}_{\mathrm{out}}=\sqrt{2\Gamma}\,\hat{\mathbf{R}}$~\cite{GardinerZoller}. 
In the Fourier space, the quadratures of input and output modes are connected via the transfer function that solves eqs.~\eqref{langevin quad}, 
$S(\omega)$, as $\hat{\mathbf{R}}_{\mathrm{out}}(\omega)=S(\omega)\hat{\mathbf{R}}_{\mathrm{in}}(\omega)$~\cite{Gouzien2020} where $\omega\in\mathbb{R}$. 
They are conjugate symmetric with respect to the transformation $\omega\leftrightarrow - \omega$, $\hat{\mathbf{R}}^ {\dag}(\omega)=\hat{\mathbf{R}}(-\omega)$, 
so as to ensure their Hermiticity in time domain~\cite{quadfourier}. $S(\omega)$ is an $\omega$-symplectic matrix-valued function of the $\omega$~\cite{omegasymp} (see Eq.~\eqref{S annx} in the Appendix). 
\\

\textbf{Morphing supermodes analysis}.
As demonstrated in~\cite{Gouzien2020}, in the general case of a system presenting both linear and nonlinear dispersion (in $G$) and parametric amplification (in $F$), 
squeezing properties need to be described in terms of \textit{morphing supermodes}. These are coherent superpositions of the original frequency modes that evolve with a continuous parameter (here $\omega$). 
They allow mapping multimode CV entangled states into a collection of $N$ independent squeezed states. 
The explicit shape of morphing supermodes is obtained by performing an analytic Bloch-Messiah decomposition (ABMD) of the transfer function $S(\omega)=U(\omega)D(\omega)V^\dagger(\omega)$. 
In this expression, $U(\omega)$ and $V(\omega)$ are unitary and $\omega$-symplectic matrix-valued functions that characterise the supermode structure. 
Correspondingly, the output quadratures of morphing supermodes read as
\begin{align}\label{morsup}
\hat{\bm{R}}'_{\mathrm{out}}(\omega)&=U^\dagger(\omega)\hat{\bm{R}}_{\mathrm{out}}(\omega).
\end{align}
These linear combinations of cavity modes change smoothly with $\omega$ but lead to the optimally \mbox{(anti-)squeezed} quadratures. 
The actual value of their noise levels is given by the elements of the diagonal matrix 
$D(\omega)= \diag\{d_1(\omega),\ldots,d_N(\omega)|\,d_1^{-1}(\omega),\ldots,d_N^{-1}(\omega)\}$, where $d_i^ {-1}(\omega)$ is the squeezing of supermode ``i" and  
$d_i(\omega)$ its anti-squeezing (with $d_i(\omega)\ge1$ for all $\omega$). Remarkably, these values correspond to the optimal \mbox{(anti-)squeezing} provided by the system. 

In the time domain, assuming input vacuum state, the stationary Gaussian quantum state at the micro-resonator output is entirely characterised by the covariance matrix 
$\sigma_{\mathrm{out}}(t)=\frac{1}{2}\langle \bm{R}_{\mathrm{out}}(0)\bm{R}_{\mathrm{out}}^{\transp}(t)
+{(\bm{R}_{\mathrm{out}}(t)\bm{R}^{\transp}_{\mathrm{out}}(0))}^{\transp}\rangle$~\cite{Kolobov2011}. In Fourier domain it corresponds to the spectral covariance matrix,
\begin{align}\label{spectralcov}
\sigma_{\mathrm{out}}(\omega)
=
\frac{1}{2\sqrt{2\pi}}U(\omega)D^2(\omega)U^{\dag}(\omega).
\end{align}
Note that, in general, $\sigma_{\mathrm{out}}(\omega)$ is hermitian since $D(\omega)$ is real.

The morphing supermode analysis allows describing in details the squeezing features of the synchronously pumped micro-resonator. 
In this regard, note that due to the extremely general form of equations describing its linearized dynamics, the analysis derived here apply to an extremely broad class of multimode gaussian states, 
all characterised by a hermitian covariance matrix as in Eq.~\eqref{spectralcov}. Discussed results can thus easily be extended to many other situations. \\

\begin{figure}[t!]
\includegraphics[width=\columnwidth]{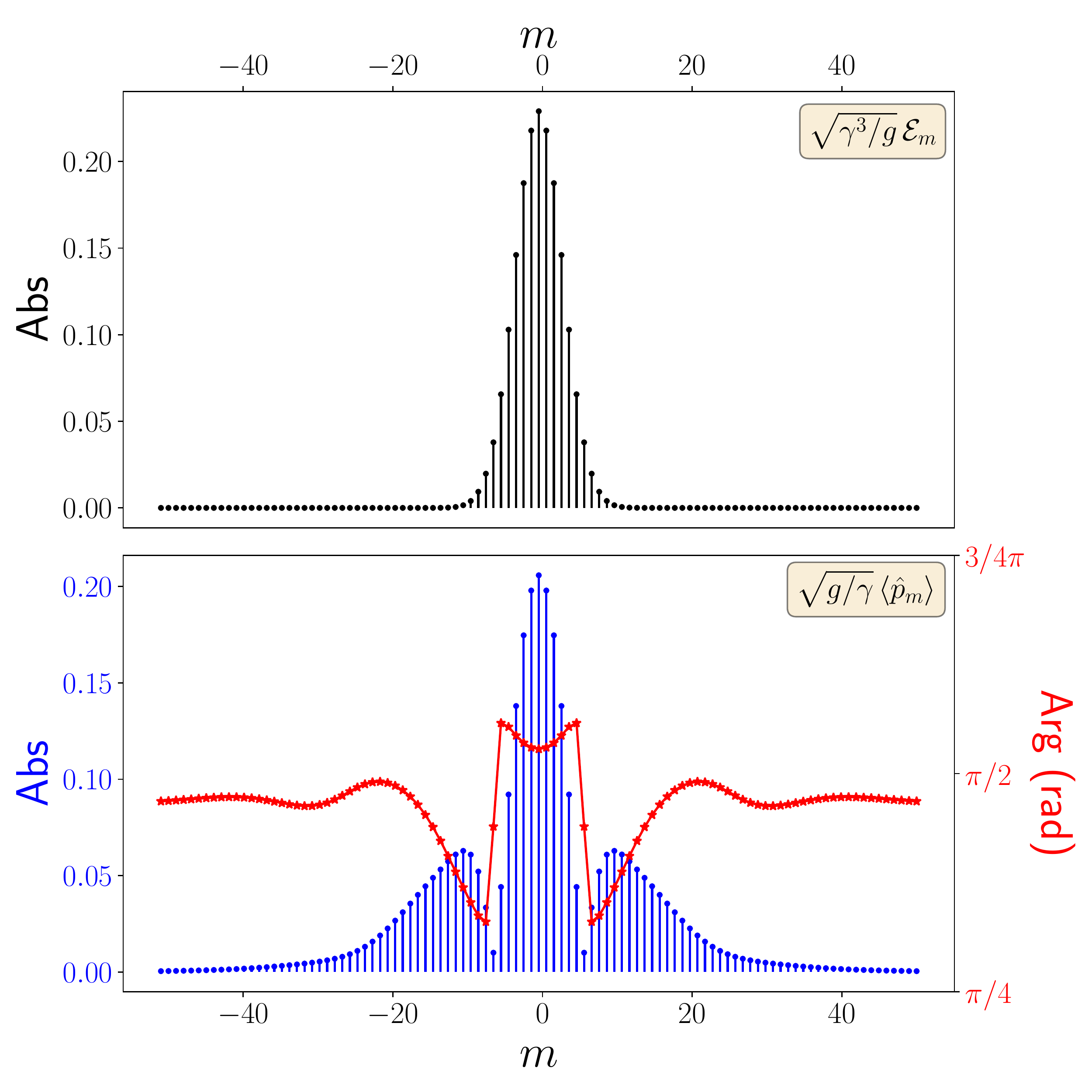}
\caption{(\emph{Top}) Normalized amplitude and phase profiles of the injection $\sqrt{\gamma^3/g}\,\mathcal{E}_m$~(see Appendices) 
and (\emph{bottom}) intra-cavity steady state solutions $\sqrt{\gamma/g}\,\langle\hat{p}_m\rangle$.
Parameters: $\mathcal{E}_m$ has a Gaussian spectral distribution $\mathcal{E}_0\exp{-m^2/(2\sigma_{\mathrm{p}})}$ with 
$\sigma_{\mathrm{p}}=20$ and $\mathcal{E}_0$ is chosen so that the system is $1\%$ below its threshold;
$\Delta_0/\gamma=0$, $\Delta\Omega/\gamma=0$, $\Omega_2/\gamma=-0.01$ and losses equal to $\gamma$ for all $m$. 
The stable steady state solution $\langle\hat{p}_m\rangle$ (Fig.~\ref{Figure2}-bottom) is obtained by solving the classical part of the nonlinear equations~\eqref{Langpump} and~\eqref{Langsignal} 
for a space of $N=101$ pump modes. }
\label{Figure2}
\end{figure}

\textbf{Multimode squeezing from a micro-resonator.}
As a representative example, it will be considered the case of a pump  frequency comb of spectral amplitudes $\mathcal{E}_m$ with Gaussian distribution
$\mathcal{E}_0\exp(-m^2/(2\sigma_{\mathrm{p}}))$,  
resonant with the central cavity mode $m=0$ \emph{i.e.}\@ $\Delta_0=0$ (Fig.~\ref{Figure2}-top). Its repetition rate matches the double of the cavity average FSR ($\Delta\Omega=0$). 
In the numerical simulations, the spectral width is $\sigma_{\mathrm{p}}=20$, and $\mathcal{E}_0$ is set so that the system is $1\%$ below its threshold. 
Cavity losses are equal $\gamma_m=\gamma$ for all $m$ and second order anomalous dispersion is set to $\Omega_2=-0.01\gamma$.

\begin{figure}[t!]
\includegraphics[width=0.8\columnwidth]{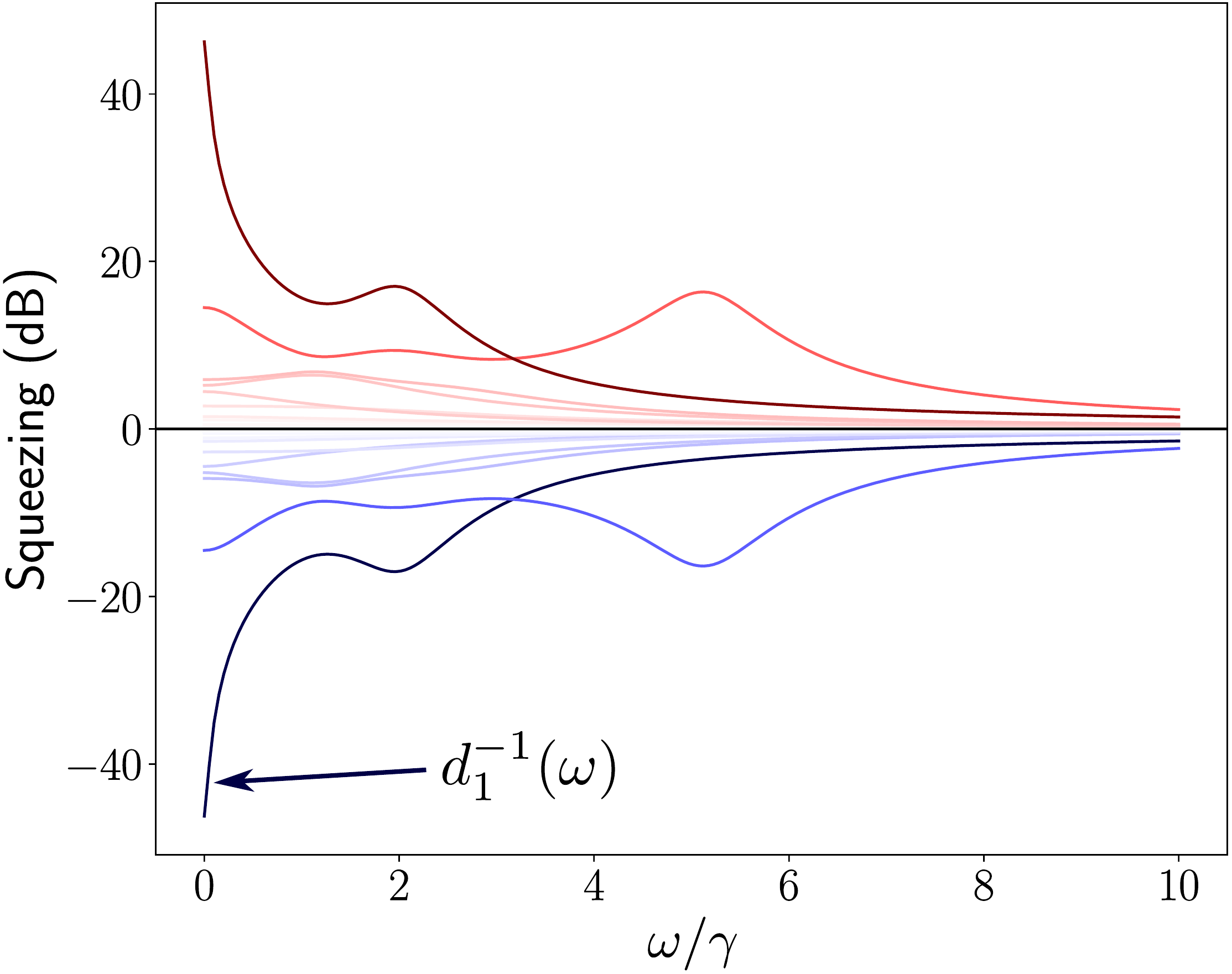}
\caption{Frequency-dependent singular values $d_i^2(\omega)$ and $d_i^{-2}(\omega)$.
They quantify the degree of anti-squeezing $d_i(\omega)$ (shades of red) and squeezing $d_i^ {-1}(\omega)$ (shades of blue) respectively.
The zero level represents the standard quantum limit. The function $d_1^{-1}(\omega)$ corresponds to the optimal level of squeezing associated to the morphing supermode
$\bm{U}_1(\omega)$ in Fig.~\ref{Figure3}.
Parameters: $\mathcal{E}_m$ has a Gaussian spectral distribution $\mathcal{E}_0\exp{-m^2/(2\sigma_{\mathrm{p}})}$ with 
$\sigma_{\mathrm{p}}=20$ and $\mathcal{E}_0$ is chosen so that the system is $1\%$ below its threshold;
$\Delta_0/\gamma=0$, $\Delta\Omega/\gamma=0$, $\Omega_2/\gamma=-0.01$ and losses equal to $\gamma$ for all $m$.}
\label{Figure:sqz}
\end{figure}

As illustrated in Fig.~\ref{Figure2}, the initially real Gaussian injection profile (top) results into a complex intracavity steady state (bottom). 
Its non-trivial amplitude and phase spectral profiles enter the systems dynamics via eqs.~\eqref{F} and~\eqref{G}. 
Correspondingly, Fig.~\ref{Figure:sqz} shows, for $i=\{1,\ldots,N\}$, optimal squeezing ($d_i^{-1}(\omega)$) and antisqueezing ($d_i(\omega)$) levels as functions of $\omega$, as obtained by ABMD. 
At $\omega=0$, the highest value of squeezing  is obtained for first supermode ($i=1$) and the highest value of anti-squeezing corresponds to the $N+1$-th supermode.
The frequency mode combination that gives the squeezed quadrature of the first morphing supermode is obtained by $\bm{U}_{1}(\omega)$, \emph{i.e.} by the first column of $U(\omega)$, 
and it is represented in Fig.~\ref{Figure3}-top. 
Similar curves are also observed for higher order supermodes~(see Appendices).
As it can be seen, the ABMD returns supermodes whose structure smoothly depends on $\omega$ and have a real and an imaginary parts both non null. 
As a consequence, the multimode quantum state produced by the micro-resonator is characterised by a spectral covariance matrix~\eqref{spectralcov} that, 
contrarily to what was assumed in previous studies, is not real. This formally reflects the presence of an imbalance between the fluctuations of the noise spectral components at $\omega$ 
and $-\omega$~\cite{Barbosa2013a,Barbosa2013b}.
Such an effect is characteristic of a dynamics in a $\chi^{(3)}$ medium and of a mode dependent dispersion. It is not present in dispersion-compensated nonlinear cavities with $\chi^{(2)}$ media whose interaction matrix leads to $G=0$ and, correspondingly, to supermodes that are frequency independent and real in the quadrature representation~\cite{Patera2010,Arzani2018}.\\
\textbf{Homodyne detection and measurable squeezing}. In experiments, the spectral covariance matrix of Eq.~\eqref{spectralcov} can be reconstructed via frequency homodyning: 
a reference beam, called ``local oscillator'' (LO), beats with the micro-resonator output and the Fourier transform of its photodetection signal is performed. 
Such a projective measurement allows retrieving the field quadratures and, in particular, the measured noise spectrum:
\begin{align}\label{measure}
\Sigma_{\bm{Q}}(\omega)
&=
\bm{Q}^\transp\sigma_{\mathrm{out}}(\omega)\bm{Q}.
\end{align}
In this context, $\omega$ is indicated as the so-called analysis frequency as it directly identifies a given noise component of the photocurrent signal. 
Its value can be experimentally set depending on the specific practical situation. In Eq.~\eqref{measure}, the normalised column vector $\bm{Q}$ corresponds 
to the spectral profile of the LO in the quadrature representation. Note that $\bm{Q}$ must be a real vector so as to guarantee that, in time domain, 
LO quadratures and their linear combinations are hermitian operators.\\
\begin{figure}[t!]
\includegraphics[width=\columnwidth]{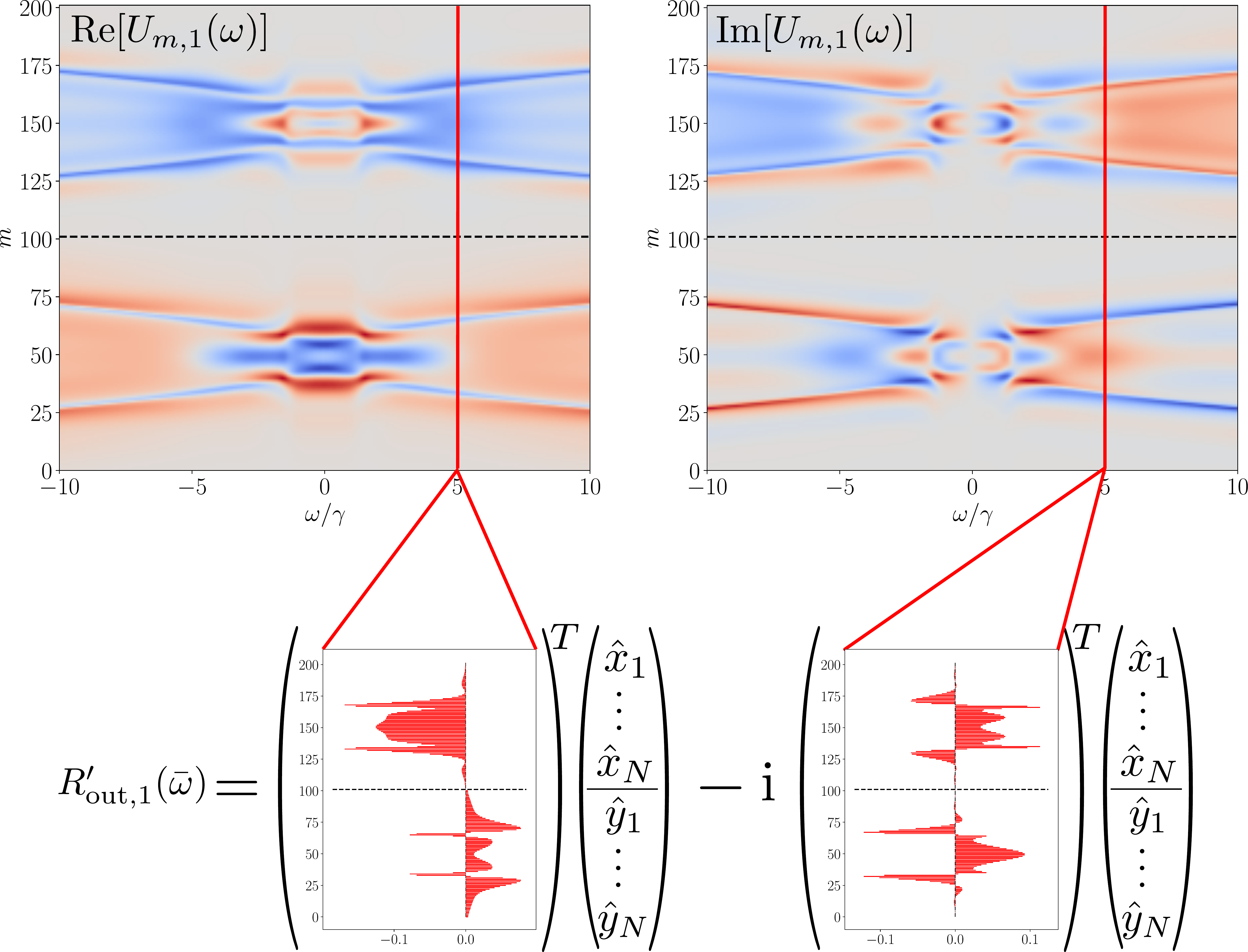}
\caption{(\emph{Top}) real and imaginary part of the first output morphing supermode for the case of a micro-resonator.
(\emph{Bottom}) for a given frequency $\bar{\omega}$, the colum vectors of the real and imaginary part of $\bm{U}_1(\omega)$ give the coefficients 
of the supermode quadrature $R'_{\mathrm{out},1}(\bar{\omega})$ according to expression~\eqref{Rprime}.
They also define the profile a LO should have in order to detect the optimal level of squeezing $d_1^{-1}(\bar{\omega})$ (see Fig.~\ref{Figure:sqz}). 
Parameters: $\mathcal{E}_m$ has a Gaussian spectral distribution $\mathcal{E}_0\exp{-m^2/(2\sigma_{\mathrm{p}})}$ with 
$\sigma_{\mathrm{p}}=20$ and $\mathcal{E}_0$ is chosen so that the system is $1\%$ below its threshold;
$\Delta_0/\gamma=0$, $\Delta\Omega/\gamma=0$, $\Omega_2/\gamma=-0.01$ and losses equal to $\gamma$ for all $m$.}
\label{Figure3}
\end{figure}
Fig.~\ref{Figure3}-bottom illustrates how to practically obtain, at a given $\bar{\omega}$, the first supermode quadrature from the matrix $U(\omega)$. From~\eqref{morsup} 
and the property $U^\dagger(\omega)=U(-\omega)$~\cite{Gouzien2020}, the quadrature operator can be expressed in terms of real and complex linear combinations
\begin{equation}
R'_{\mathrm{out},1}(\bar{\omega})=\big(\mathrm{Re}[\bm{U}_{1}^\transp(\bar{\omega})]-\mathrm{i}\,\mathrm{Im}[\bm{U}_{1}^\transp(\bar{\omega})]\big)\bm{R}_{\mathrm{out}}(\omega).
\label{Rprime}
\end{equation}
Changing the analysis frequency thus implies changing the linear combination. 
Such a morphing behaviour has a strong practical impact on the way squeezing outside the micro-resonator should be experimentally measured. 
By inserting eq.~\eqref{spectralcov} in~\eqref{measure}, it is evident that optimal squeezing $d_i^{-1}(\omega)$ (anti-squeezing $d_i(\omega)$) can be measured only if $\bm{Q}$ matches the $i$-th column of 
$U(\omega)$ for all $\omega$ (\emph{i.e.}\@ $U(\omega)$ projects optimally on the LO). 
However, in general, this is not possible for two reasons: (i) $\bm{Q}$ should depend on $\omega$ and (ii) $\bm{Q}$ is real while $U(\omega)$ can be complex. 
In the case $U(\omega)$ is real and $\bm{Q}$ constant, the homodyne detection can detect optimal squeezing only at the frequency $\bar{\omega}$ 
for which the local oscillator matches the supermode profile ($\bm{Q}^\transp\, \bm{U}_{1}(\bar{\omega})=1$). Reconstructing the squeezing profile demands being able 
to reshape $\bm{Q}$ for each choice of $\bar{\omega}$. On the other hand, since in general $U(\omega)$ is not real and the spectral profile of the LO can only be real, 
homodyne detection can measure only the real part of the supermode quadratures. In other words, when $U(\omega)$ is complex the homodyne measure is suboptimal for all values 
of $\omega$ and part of the quantum properties of the output state remains hidden.
To retrieve the optimal squeezing, the LO profile should be a complex-valued smooth function of $\omega$. 
As discussed in~\cite{Gouzien2020}, this cannot be implemented with a standard detection scheme and rather requires an interferometer with memory effect. 
The description of such a device is beyond the scope of this work and will be the subject of a subsequent publication.\\
\begin{figure*}[t!]
\includegraphics[width=\textwidth]{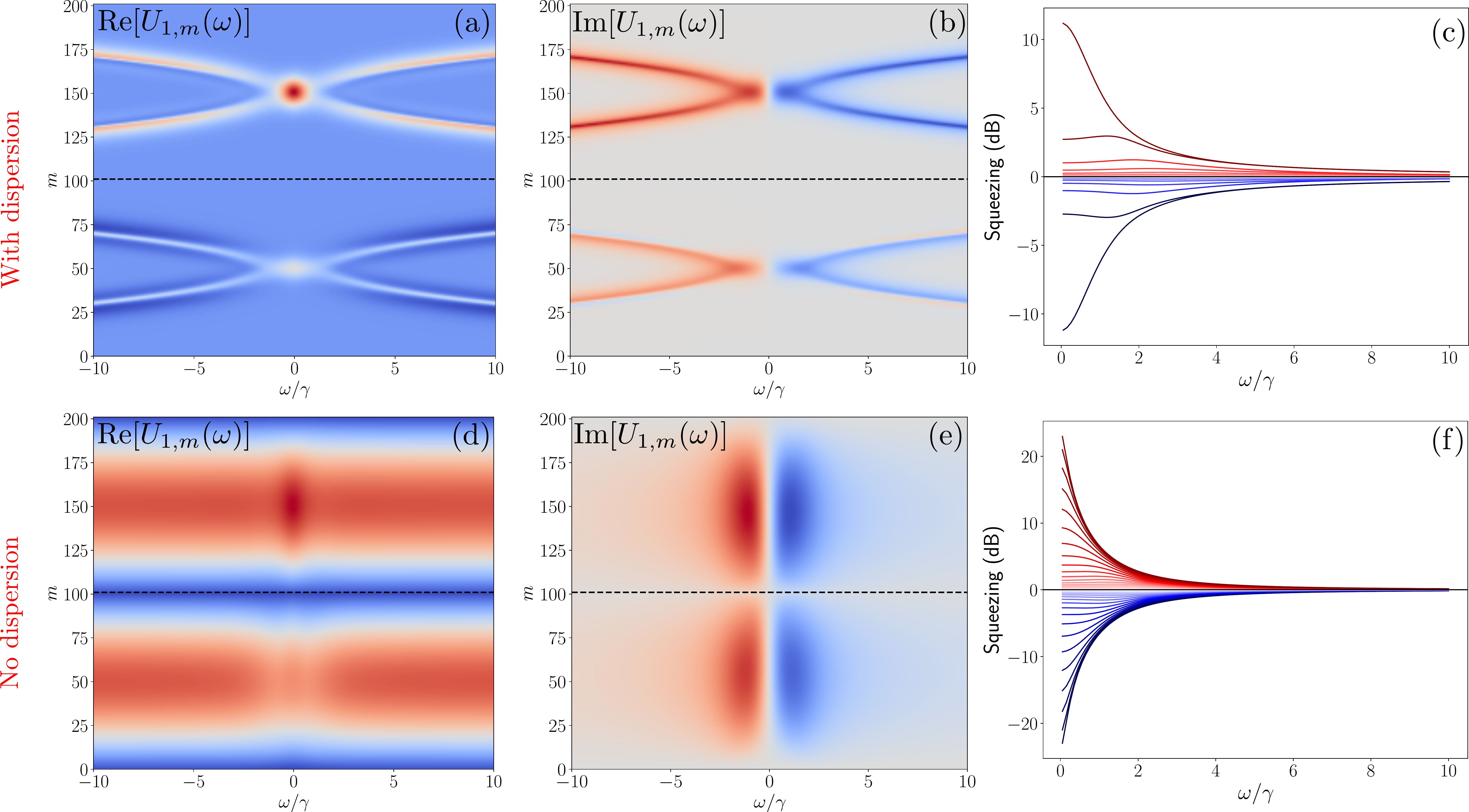}
\caption{Case of Gaussian intracavity steady state solution $\langle\hat{p}_m\rangle=A\exp(-m^ 2/2\sigma)$:
real (a) (resp. (d)) and imaginary (b) (resp. (e)) part of the first morphing supermode and spectrum of the frequency-dependent singular values (c) (resp. (f)) $d_i^2(\omega)$ and $d_i^{-2}(\omega)$
in the case of micro-ring without dispersion (resp. with dispersion).
Parameters: $A=0.12$, $\sigma=10$, $\Delta_0/\gamma=-2$, $\Delta\Omega/\gamma=0$, and (a)-(c) $\Omega_2/\gamma=-0.01$, (d)-(f) $\Omega_2/\gamma=0$, 
with losses equal to $\gamma$ for all $m$.}
\label{Figure:engineered}
\end{figure*}
Complex morphing supermodes are obtained for a vast majority of sets of micro-resonator parameters. 
From a physical point of view, this behaviour is to be associated with the presence of self- and cross-phase modulation due to FWM and of dispersion-dependent detuning (both included in matrix $G$), 
that scramble the quantum correlations generated through the parametric amplification processes (in $F$). 
These combined factors heavily affects the pump intracavity steady states $\langle\hat{p}_m\rangle$, leading to strong deformation of both its amplitude and phase profiles (see Fig.~\ref{Figure2}).
As a consequence, it is pertinent to consider the case of a pump beam whose spectrum has been engineered before the micro-resonator as a possible strategy 
to make squeezing detectable with a standard homodyne measurement. 
Figure~\ref{Figure:engineered}-top shows the first morphing supermode as obtained when the input pump profile $\{\mathcal{E}_m\}$ is engineered 
so as to obtain a Gaussian real intracavity steady state centred in $m=0$ as a solution of the classical part of the non-linear Langevin equations (eqs.~\eqref{Langpump} in Appendix), 
\emph{i.e.}\@ $\langle\hat{p}_m\rangle=A\exp(-m^ 2/2\sigma)$. All the other parameters are kept as in the previous case.
Although less complicated, the first morphing supermode shows a non trivial imaginary part, a simpler frequency dependence but 
a somehow reduced level of squeezing due to an increased distance to the threshold. 
Better results are obtained when considering, in addition to the pump engineering, a microring with negligible dispersion ($\Omega_2=0$). 
In experiment, such a condition can be implemented thanks to specially tailored waveguide geometries~\cite{Dirani}.
As shown in Figure~\ref{Figure:engineered}-bottom, in this case, the morphing behaviour is strongly amended and, remarkably, 
the supermode profile is weakly depending on $\omega$. This makes the real part of the covariance matrix detectable with a standard homodyne measurement. \\

\textbf{Conclusions}. 
This work provides a complete characterisation of the multimode quantum properties of silicon-based resonators operating in pulsed regime (synchronously pumped). 
The analysis is done in terms of squeezed morphing supermodes~\cite{Gouzien2020}. 
The treatment shows that, since the spectral profiles of supermodes are in general a complex function of the analysis frequency $\omega$, 
a full experimental characterisation of their quantum properties is beyond the possibility of standard homodyne detection. 
This behaviour, enlightened here for the first time, shows the need for carefully engineered experimental configurations, 
including the spectral profile of the pump and of the resonator itself (dispersion),
conceived to obtain supermodes with a weak dependence on $\omega$. A further development of the presented supermode investigation is to identify, 
on the base of a specific micro-resonator architecture and optical properties, working conditions under which the covariance matrix is real and thus fully detectable.\bigskip

\section*{Funding and Acknowledgments}
We acknoledge frutiful discussions about analytic decompositions with Alessandro Pugliese and about spectral covariance matrix with Carlos Navarrete-Benlloch. 
This work has been conducted within the framework of the project SPHIFA (ANR-20-CE47-0012).  

Virginia D'Auria acknowledges financial support from the Institut Universitaire de France (IUF).
\FloatBarrier
\section*{Appendices}\label{Appendices}
\subsection{The Hamiltonian}\label{AppendixA}
In order to establish the system Hamiltonian, first we identify the pertinent cavity resonances that are involved in the FWM process.
Under the hypothesis that central frequency of the injection $\omega_{\mathrm{p}}$ is close to the cavity resonance $\omega_{m_0}$, 
the pump will excite one cavity resonance every two as depicted in figure~\ref{Figure1}. 
Thus we can define the injection reference comb of frequencies
\begin{equation}
\bar{\omega}_m=\omega_{\mathrm{p}}+\frac{\Omega_{\mathrm{p}}}{2}(m-m_0)
\end{equation}
and, without loss of generality, we can set $m_0=0$.
Among these frequencies, we can distinguish the reference pump comb $\bar{\omega}_{\mathrm{p},m}$ and the reference signal comb $\bar{\omega}_{\mathrm{s},m}$ such that
\begin{align}
\bar{\omega}_{\mathrm{p,m}}&=\omega_{\mathrm{p}}+2m\frac{\Omega_{\mathrm{p}}}{2},
\label{refpump}
\\
\bar{\omega}_{\mathrm{s,m}}&=\omega_{\mathrm{p}}+(2m+1)
\frac{\Omega_{\mathrm{p}}}{2}.
\label{refsignal}
\end{align}
We label with $\omega_{\mathrm{p},m}$ ($\omega_{\mathrm{s},m}$) the cavity resonances $\omega_m$ (see eq.~\eqref{resonanceTaylor}) of even (odd) order,
that are the closest to the pump reference frequencies $\bar{\omega}_{\mathrm{p,m}}$ (signal reference frequencies $\bar{\omega}_{\mathrm{s,m}}$).

Field quantification is performed with respect to the quadratic part of the system Hamiltonian and later we will treat the higher order terms as a perturbation.
In order to correctly keep into account dispersion we chose the displacement field $\hat{\mathbf{D}}$ and 
the magnetic field $\hat{\mathbf{B}}$ as the fundamental entities~\cite{Drummond1999,Quesada2017,Raymer2020} (bold designate vector quantities). 
Then, in order to distinguish the intracavity modes that are populated by the external pump and those that are not, we decompose the displacement field as
\begin{equation}
\hat{\mathbf{D}}(\mathbf{r},t)=
\left[
\hat{\mathbf{D}}_{\mathrm{p}}(\mathbf{r},t)+\hat{\mathbf{D}}_{\mathrm{s}}(\mathbf{r},t)\right]
\end{equation}
where, in the Schr\"{o}dinger picture,
\begin{align}
\hat{D}_{\mathrm{p}}(\mathbf{r},t)&=\mathrm{i}\sum_m\mathcal{D}_{\mathrm{p},m}
\Big(
\hat{p}_m\mathbf{d}_{\mathrm{p},m}(\mathbf{r})
+
\hat{p}_m^{\dag}\mathbf{d}_{\mathrm{p},m}^*(\mathbf{r})
\Big),
\label{Ep}
\\
\hat{D}_{\mathrm{s}}(\mathbf{r},t)&=
\mathrm{i}\sum_m\mathcal{D}_{\mathrm{s},m}
\Big(
\hat{s}_m\mathbf{d}_{\mathrm{s},m}(\mathbf{r})
+
\hat{s}_m^{\dag}\mathbf{d}_{\mathrm{s},m}^*(\mathbf{r})
\Big)
\label{Es}.
\end{align}
The spatial modes $\mathbf{d}_{\mathrm{f},m}(\mathbf{r})$ (with $\mathrm{f}\in\{\mathrm{p},\mathrm{s}\}$) are found by solving the following equations
\begin{align}
&\nabla\wedge\left(\frac{1}{n^2(\mathbf{r},\omega_m)}\nabla\wedge\mathbf{b}_{\mathbf{f},m}(\mathbf{r})\right)=\frac{\omega_m^ 2}{c^ 2}\mathbf{b}_{\mathbf{f},m}(\mathbf{r})
\\
&\mathbf{d}_{\mathrm{f},m}(\mathbf{r})=
\frac{\mathrm{i}\,c}{\omega_{\mathrm{f},m}}\nabla\wedge\mathbf{b}_{\mathrm{f},m}(\mathbf{r})
\end{align}
they are normalized such as
\begin{align}
\int\mathrm{d}^3\mathbf{r}
\frac{\mathbf{d}_{\mathrm{f},m}^*(\mathbf{r})\cdot\mathbf{d}_{\mathrm{f},m}(\mathbf{r})}{\epsilon_0 n^2(\mathbf{r},\omega_{\mathrm{f},m})}
\frac{v_{\phi}(\omega_{\mathrm{f},m})}{v_{g}(\omega_{\mathrm{f},m})}
=1
\end{align}
where $v_{\phi}(\omega_{\mathrm{f},m})$ and $v_{g}(\omega_{\mathrm{f},m})$ are the phase and group velocities, respectively.
The operators $\hat{p}_m(t)$ and $\hat{s}_m(t)$ are the slowly-varying annihilation field amplitudes for pump and signal fields. 
They destroy one elemental excitation in the pump (respectively signal) mode $\mathbf{d}_{\mathrm{p},m}(\mathbf{r})$ (resp. $\mathbf{d}_{\mathrm{s},m}(\mathbf{r})$) 
and verify the standard boson commutation rules
\begin{align}
\left[\hat{p}_m,\hat{p}_n^{\dag}\right]&=\delta_{m,n},
\\
\left[\hat{p}_m,\hat{p}_n\right]&=0,
\\
\left[\hat{s}_m,\hat{s}_n^{\dag}\right]&=\delta_{m,n},
\\
\left[\hat{s}_m,\hat{s}_n\right]&=0.
\end{align}
The quantities $\mathcal{D}_{\mathrm{f},m}$ (with $\mathrm{f}\in\{\mathrm{p},\mathrm{s}\}$) are given by
\begin{equation}
\mathcal{D}_{\mathrm{f},m}=\sqrt{\frac{\epsilon_0\hbar\omega_{\mathrm{f},m}}{2}}
\end{equation}
and can be interpreted as the single polariton field amplitudes in the mode $\mathbf{d}_{\mathrm{f},m}(\mathbf{r})$. 
They have the form $\mathbf{d}_{\mathrm{f},m}(\mathbf{r})=R(r,z)Y_{\mathrm{f},m}(\theta)\mathbf{u}$ and $\mathbf{u}\approx\mathbf{u}_{r}$.

Since we are in the context of a scalar theory, the nonlinear polarization is also along the radial vector $\mathbf{u}_r$ and its component takes the form
\begin{equation}
\hat{P}_{\mathrm{nl}}(\mathbf{r},t)=-\epsilon_0\eta^{(3)}D^3(\mathbf{r},t),
\end{equation}
where $\eta^{(3)}$ is the inverse permittivity tensor and assuming a medium with null second order susceptibility. 
The nonlinear interaction Hamiltonian is then  
\begin{equation}
\hat{H}_{\mathrm{int}}=
\frac{\eta^{(3)}}{4}\int\hat{D}^4\,\mathrm{d}^3\mathbf{r}.
\label{NLint}
\end{equation}
The application of the rotating-wave approximation to~\eqref{NLint}, after using expressions eqs.~\eqref{Ep} and~\eqref{Es}, 
allows to keep only three kind of processes (and their reciprocal) that conserve the energy: the first process converts two pump photons 
into two other pump photons such that $\omega_{\mathrm{p},m}+\omega_{\mathrm{p},n}=\omega_{\mathrm{p},l}+\omega_{\mathrm{p},k}$; 
the second converts two pump photons into two signal/idler photons such that $\omega_{\mathrm{p},m}+\omega_{\mathrm{p},n}=\omega_{\mathrm{s},l}+\omega_{\mathrm{s},k}$; 
the third converts one pump photon and one signal/idler photon to another couple of pump and signal/idler photons such that 
$\omega_{\mathrm{p},m}+\omega_{\mathrm{s},n}=\omega_{\mathrm{p},l}+\omega_{\mathrm{s},k}$. The processes 
$\omega_{\mathrm{s},m}+\omega_{\mathrm{s},n}=\omega_{\mathrm{s},l}+\omega_{\mathrm{s},k}$ are neglected because, in the semi-classical approximation, they are mediated
by amplitudes that have null mean value ($\langle\hat{s}_m\rangle$=0).
As a consequence the nonlinear interaction term takes the form
\begin{align}
\hat{H}_{\mathrm{int}}\approx&
\hat{H}_{\mathrm{p,p}}+\hat{H}_{\mathrm{p,s}},
\end{align}
with
\begin{align}
\hat{H}_{\mathrm{p,p}}=&
\frac{g_0}{6}\sum_{k,l,m,n}
\mathcal{A}_{k,l}^{m,n}
\Big(
\delta_{[k+l-m-n]}\,
\hat{p}_k\hat{p}_l\hat{p}_m^{\dag}\hat{p}_n^{\dag}+
\nonumber\\
&+
\delta_{[k-l+m-n]}\,
\hat{p}_k\hat{p}_l^{\dag}\hat{p}_m\hat{p}_n^{\dag}+
\delta_{[k-l-m+n]}\,
\hat{p}_k\hat{p}_l^{\dag}\hat{p}_m^{\dag}\hat{p}_n
\Big)
+
\nonumber\\
&+\mathrm{H.c.},
\end{align}
and
\begin{align}
\hat{H}_{\mathrm{p,s}}=&
g_0\sum_{k,l,m,n}\mathcal{B}_{k,l}^{m,n}
\Big(
\delta_{[k+l-m-n-1]}\,
\hat{p}_k\hat{p}_l\hat{s}_m^{\dag}\hat{s}_n^{\dag}
+
\nonumber\\
&+
\delta_{[k-l+m-n]}\,
\hat{p}_k\hat{p}_l^{\dag}\hat{s}_m\hat{s}_n^{\dag}+
\delta_{[k-l-m+n]}\,
\hat{p}_k\hat{p}_l^{\dag}\hat{s}_m^{\dag}\hat{s}_n
\Big)
+
\nonumber\\
&+
\mathrm{H.c.}
\end{align}
where ``H.c." stands for ``Hermitian conjugate".
In these expressions, $\delta_{[\cdot]}$ is the usual Kronecker symbol (equal to 1 when $[\cdot]=0$ and to 0 otherwise), $g_0$ is the nonlinear coupling constant
\begin{equation}
g_0=\frac{3\hbar^2\epsilon_0^2\eta^{(3)}\Lambda}{8}
\end{equation}
and
\begin{align}
\Lambda&=
\int_{0}^{+\infty}\int_{-\infty}^{+\infty}
\mathrm{d}r\,\mathrm{d}z\,
\,r\left|R(r,z)\right|^4,
\\
\mathcal{A}_{k,l}^{m,n}&=
\sqrt{\omega_{\mathrm{p},k}\omega_{\mathrm{p},l}\omega_{\mathrm{p},m}\omega_{\mathrm{p},n}},
\\
\mathcal{B}_{k,l}^{m,n}&=
\sqrt{\omega_{\mathrm{p},k}\omega_{\mathrm{p},l}\omega_{\mathrm{s},m}\omega_{\mathrm{s},n}}.
\end{align}
In the following we will assume, for all $k,l,m,n$, $\mathcal{A}_{k,l}^{m,n}\approx\omega_0^2$ and 
$\mathcal{B}_{k,l}^{m,n}\approx\omega_0^2$.

The system total Hamiltonian is then
\begin{equation}
H_{\mathrm{tot}}=H_0+H_{\mathrm{int}}+H_{\mathrm{inj}}
\end{equation}
where
\begin{equation}
\hat{H}_0=
\sum_{m}
\hbar\omega_{\mathrm{p},m}\hat{p}_m^{\dag}\hat{p}_m +
\sum_{m}
\hbar\omega_{\mathrm{s},m}\hat{s}_m^{\dag}\hat{s}_m
\end{equation}
is the Hamiltonian of the free fields and
\begin{equation}
\hat{H}_{\mathrm{inj}}=
\mathrm{i}\hbar
\sum_{m}
\Big(
\mathcal{E}_m \hat{p}_m^{\dag}\mathrm{e}^{-\mathrm{i}\bar{\omega}_{\mathrm{p},m}t}+
\mathcal{E}_m^* \hat{p}_m \mathrm{e}^{\mathrm{i}\bar{\omega}_{\mathrm{p},m}t}
\Big)
\label{Hinj}
\end{equation}
describes the injection of a frequency comb (synchronous pumping) with spectral amplitudes $\mathcal{E}_m$ at frequencies 
$\bar{\omega}_{\mathrm{p},m}$.
\subsection{Multimode quantum Langevin equations}\label{MultiMulti}

The Heisenberg equations for pump and signal fields are:
\begin{align}
\mathrm{i}\hbar\frac{\mathrm{d}\hat{p}_j}{\mathrm{d}t}=&
\hbar\omega_{\mathrm{p},j}\hat{p}_j
+
\mathrm{i}\hbar\mathcal{E}_j\mathrm{e}^{-\mathrm{i}\bar{\omega}_{\mathrm{p},j}t}
+
\nonumber\\
&+
\frac{2g_0\omega_0^2}{3}\sum_{m,n}
\Big(
\hat{p}_{m+n-j}^\dag\hat{p}_m\hat{p}_n+
\hat{p}_{m+n-j+1}^\dag\hat{s}_m\hat{s}_n
+
\nonumber\\
&+
\hat{p}_{j-m+n}\left(
\hat{p}_m\hat{p}_n^\dag+
\hat{p}_n^\dag\hat{p}_m+
\hat{s}_m\hat{s}_n^\dag+\hat{s}_n^\dag\hat{s}_m
\right)
\Big),
\label{Heipump}
\\
\mathrm{i}\hbar\frac{\mathrm{d}\hat{s}_j}{\mathrm{d}t}=&
\hbar\omega_{\mathrm{s},j}\hat{s}_j
+
2g_0\omega_0^2
\sum_{m,n}
\Big(
\hat{p}_{j-m+n+1}\hat{p}_m\hat{s}_n^\dag
+
\nonumber\\
&+
\hat{p}_{j+m-n}\hat{p}_m^\dag\hat{s}_n
+
\hat{p}_{m+n-j}^\dag\hat{p}_m\hat{s}_n
\Big)
\label{Heisignal}
\end{align}
The explicit time dependence in~\eqref{Heipump} can be removed by moving to the reference frame of the injection. Hence we define new fields such that
\begin{align}
\hat{p}_m&\rightarrow\hat{p}_m\mathrm{e}^{-\mathrm{i}\bar{\omega}_{\mathrm{p},m}t},
\\
\hat{s}_m&\rightarrow\hat{s}_m\mathrm{e}^{-\mathrm{i}\bar{\omega}_{\mathrm{s},m}t}
\end{align}
and write eqs.~\eqref{Heipump} and~\eqref{Heisignal} as
\begin{align}
\frac{\mathrm{d}\hat{p}_j}{\mathrm{d}t}=&
-\mathrm{i}\Delta_{\mathrm{p},j}\hat{p}_j+\mathcal{E}_j+
\nonumber\\
&-\mathrm{i}
\frac{g}{3}
\sum_{m,n}
\Big(
\hat{p}_{m+n-j}^\dag\hat{p}_m\hat{p}_n+
\hat{p}_{m+n-j+1}^\dag\hat{s}_m\hat{s}_n
+
\nonumber\\
&+
\hat{p}_{j-m+n}\left(
\hat{p}_m\hat{p}_n^\dag+
\hat{p}_n^\dag\hat{p}_m+
\hat{s}_m\hat{s}_n^\dag+\hat{s}_n^\dag\hat{s}_m
\right)
\Big),
\label{Heipump2}
\\
\frac{\mathrm{d}\hat{s}_j}{\mathrm{d}t}=&
-\mathrm{i}\Delta_{\mathrm{s},j}\hat{s}_j
-\mathrm{i}
g
\sum_{m,n}
\Big(
\hat{p}_{j-m+n+1}\hat{p}_m\hat{s}_n^\dag
+
\nonumber\\
&+
\Big(
\hat{p}_{j+m-n}\hat{p}_m^\dag
+
\hat{p}_{m+n-j}^\dag\hat{p}_m
\big)\hat{s}_n
\Big)
\label{Heisignal2}
\end{align}
with $g=(2g_0\omega_0^2)/(\hbar)$, $\Delta_{\mathrm{p},j}=\omega_{\mathrm{p},j}-\bar{\omega}_{\mathrm{p},j}$ and $\Delta_{\mathrm{s},j}=\omega_{\mathrm{s},j}-\bar{\omega}_{\mathrm{s},j}$. 
They are frequency dependent detunings that, after using eq.~\eqref{resonanceTaylor}, can be expressed as
\begin{align}
\Delta_{\mathrm{p},j}&\approx\Delta_0+\Delta\Omega\, (2j)+\frac{\Omega_2}{2!}(2j)^2,
\\
\Delta_{\mathrm{s},j}&\approx\Delta_0+\Delta\Omega\, (2j+1)+\frac{\Omega_2}{2!}(2j+1)^2,
\end{align}
where $\Delta_0=\omega_{0}-\bar{\omega}_{\mathrm{p}}$ is the detuning between the central cavity resonance (of order $j=0$) 
and the external injection centered at frequency $\omega_{\mathrm{p}}$, $\Delta\Omega=\Omega_1-\Omega_{\mathrm{p}}/2$ is the 
mismatch between the average FSR and the half of the spacing of the external frequency comb.
Langevin equations also include the effect of propagation losses inside the microring, that couples the pump and signal modes 
with the input vacuum modes $\hat{q}_{\mathrm{in},m}$ and $\hat{r}_{\mathrm{in},m}$ via the coefficients $\kappa_{\mathrm{p},m}$ and $\kappa_{\mathrm{s},m}$, respectively. 
In a similar way, losses due to the microring coupling with the straight guide introduce $\hat{p}_{\mathrm{in},j}$ and $\hat{s}_{\mathrm{in},m}$ via the 
coefficients $\gamma_{\mathrm{p},m}$ and $\gamma_{\mathrm{s},m}$. The explicit expression of the quantum Langevin equations reads as~\cite{GardinerZoller}:
\begin{align}
\frac{\mathrm{d}\hat{p}_j}{\mathrm{d}t}=&
-\big(\gamma_{\mathrm{p},j}+\kappa_{\mathrm{p},j}+\mathrm{i}\Delta_{\mathrm{p},j}\big)\hat{p}_j+\mathcal{E}_j+
\nonumber\\
&+\sqrt{2\gamma_{\mathrm{p},j}}\hat{p}_{\mathrm{in},j}
+\sqrt{2\kappa_{\mathrm{p},j}}\hat{q}_{\mathrm{in},j}+
\nonumber\\
&-\mathrm{i}
\frac{g}{3}
\sum_{m,n}
\Big(
\hat{p}_{m+n-j}^\dag\hat{p}_m\hat{p}_n+
\hat{p}_{m+n-j+1}^\dag\hat{s}_m\hat{s}_n
+
\nonumber\\
&+
\hat{p}_{j-m+n}\left(
\hat{p}_m\hat{p}_n^\dag+
\hat{p}_n^\dag\hat{p}_m+
\hat{s}_m\hat{s}_n^\dag+\hat{s}_n^\dag\hat{s}_m
\right)
\Big),
\label{Langpump}
\\
\frac{\mathrm{d}\hat{s}_j}{\mathrm{d}t}=&
-\big(\gamma_{\mathrm{s},j}+\kappa_{\mathrm{s},j}+\mathrm{i}\Delta_{\mathrm{s},j}\big)\hat{s}_j+
\nonumber\\
&+\sqrt{2\gamma_{\mathrm{s},j}}\hat{s}_{\mathrm{in},j}
+\sqrt{2\kappa_{\mathrm{s},j}}\hat{r}_{\mathrm{in},j}+
\nonumber\\
&-\mathrm{i}
g
\sum_{m,n}
\Big(
\hat{p}_{j-m+n+1}\hat{p}_m\hat{s}_n^\dag
+
\nonumber\\
&+
\big(
\hat{p}_{j+m-n}\hat{p}_m^\dag
+
\hat{p}_{m+n-j}^\dag\hat{p}_m
\big)\hat{s}_n
\Big).
\label{Langsignal}
\end{align}
%
\subsection{Linearized quantum Langevin equations}\label{linearization}
Quantum Langevin equations equations~\eqref{Langpump} and~\eqref{Langsignal} are, now, linearized around the system stable steady state solutions, 
$\langle\hat{p}_m\rangle$ and $\langle\hat{s}_m\rangle$. This work focuses on the below threshold regime 
where the steady state solutions for the signal exhibit null mean values, therfore we set $\langle\hat{s}_m\rangle=0, \forall m$. 
On the other hand, the $\langle\hat{p}_m\rangle$ are found as solutions of the system of nonlinear (cubic) algebraic equations obtained from the classical part of eq.~\eqref{Langpump}. 
This operation leads to a set of linear quantum Langevin equations for the signal modes expressed in terms of the quadrature column vector
$\hat{\mathbf{R}}(t)=(\hat{\mathbf{x}}(t)|\hat{\mathbf{y}}(t))^\transp$:
\begin{align}
\frac{\mathrm{d}\hat{\mathbf{R}}(t)}{\mathrm{d}t}&=
(-\Gamma'- \mathcal{K}+\mathcal{M})\hat{\mathbf{R}}(t)+
\nonumber\\
&+
\sqrt{2\Gamma'}\,\hat{\mathbf{R}}^{(\gamma)}_{\mathrm{in}}(t)+\sqrt{2\mathcal{K}}\,\hat{\mathbf{R}}^{(\kappa)}_{\mathrm{in}}(t)
\label{langevin quad double bus}
\end{align}
where the matrices $\Gamma'=\mathrm{diag}\{\gamma|\gamma\}$ and  $\mathcal{K}=\mathrm{diag}\{\kappa|\kappa\}$ are diagonal matrices containing 
the mode-dependent cavity losses due to the microring coupling  $\gamma=\mathrm{diag}\{\ldots,\gamma_{\mathrm{s},-1},\gamma_{\mathrm{s},0},\gamma_{\mathrm{s},1},\ldots\}$ 
and propagation losses $\kappa=\mathrm{diag}\{\ldots,\kappa_{\mathrm{s},-1},\kappa_{\mathrm{s},0},\kappa_{\mathrm{s},1},\ldots\}$.
The input mode quadratures are collected in the column vectors 
$\hat{\mathbf{R}}^{(\gamma)}_{\mathrm{in}}(t)=(\ldots,\hat{s}_{\mathrm{in},-1},\hat{s}_{\mathrm{in},0},\hat{s}_{\mathrm{in},+1}\ldots)^\transp$ and 
$\hat{\mathbf{R}}^{(\kappa)}_{\mathrm{in}}(t)=(\ldots,\hat{r}_{\mathrm{in},-1},\hat{r}_{\mathrm{in},0},\hat{r}_{\mathrm{in},+1}\ldots)^\transp$ and 
we suppose they are both in vacuum state. The intermodal coupling matrix $\mathcal{M}$ can be expressed as
\begin{equation}
\mathcal{M}=
\left(
\begin{array}{c|c}
\mathrm{Im}\left[G+F\right] & \mathrm{Re}\left[G-F\right]
\\
\hline
-\mathrm{Re}\left[G+F\right] & -\mathrm{Im}\left[G+F\right]^\transp
\end{array}
\right),
\label{eMMe appx}
\end{equation}
where the matrices $G$ and $F$ are such that
\begin{align}
F_{j,n}=&g
\sum_m
\langle\hat{p}_{j-m+n+1}\rangle\langle\hat{p}_m\rangle,
\label{F annx}
\\
G_{j,n}=&
\Delta_{\mathrm{s},j}\delta_{[j-n]}
+
g
\sum_m
2\langle\hat{p}_{j+m-n}\rangle\langle\hat{p}_m\rangle^*.
\label{G annx}
\end{align}
Hence $G=G^ \dag$ is an Hermitian complex matrix and $F=F^ \transp$ is symmetric. These properties make $\mathcal{M}$ an Hamiltonian matrix, that is $(\Omega \mathcal{M})^\transp = \Omega \mathcal{M}$,
with $\Omega$ the symplectic form.
%
\subsection{From double-bus to single-bus cavity Langeving equations}\label{doublebus}

In order to apply the theory we developed in~\cite{Gouzien2020}, we map eqs. \eqref{langevin quad double bus} to the linear quantum Langevin equation of a single-bus cavity.
This is obtained by defining~\cite{TheseElie}
\begin{align}
\hat{\mathbf{R}}_{\mathrm{in}}(t)
&=
\frac{
\sqrt{2\Gamma'} \hat{\mathbf{R}}^{(\gamma)}_{\mathrm{in}}(t)
+
\sqrt{2\mathcal{K}} \hat{\mathbf{R}}^{(\kappa)}_{\mathrm{in}}(t)
}
{\sqrt{2(\Gamma'+\mathcal{K})}}.
\label{Rin}
\end{align}

Hence we get the quantum Langevin equation considered in the main text, eq.~\eqref{langevin quad}
\begin{align}
\frac{\mathrm{d}\hat{\mathbf{R}}(t)}{\mathrm{d}t}&=
(-\Gamma+\mathcal{M})\hat{\mathbf{R}}(t)+
\sqrt{2\Gamma}\,\hat{\mathbf{R}}_{\mathrm{in}}(t)
\label{langevin quad annx}
\end{align}
with $\Gamma=\Gamma'+\mathcal{K}$. Then, by using the input-output relation $\hat{\mathbf{R}}_{\mathrm{out}}^{(\gamma)}=\sqrt{\Gamma'}\hat{\mathbf{R}}-\hat{\mathbf{R}}_{\mathrm{in}}^{(\gamma)}$,
the field quadratures at the output coupler $\hat{\mathbf{R}}^{(\gamma)}_{\mathrm{in}}(t)$ are given by
\begin{align}
\hat{\mathbf{R}}^{(\gamma)}_{\mathrm{out}}&=
\sqrt{\frac{\Gamma'}{\Gamma}}\hat{\mathbf{R}}_{\mathrm{out}}
+ 
\sqrt{\frac{\Gamma' \mathcal{K}}{\Gamma^2}}\hat{\mathbf{R}}^{(\kappa)}_{\mathrm{in}}
-
\left(1-\frac{\Gamma'}{\Gamma}\right)\hat{\mathbf{R}}^{(\gamma)}_{\mathrm{in}},
\end{align}
were the definition of $\hat{\mathbf{R}}_{\mathrm{out}}$ is given by~\eqref{Rin} after replacing ``in'' by ``out'' everywhere.
This column vector contains the quadratures of field operators at the output of a virtual system having only one source of losses (single-bus model).
\subsection{The omega-symplectic transfer function}\label{Somega}
The solution of the linear quantum Langevin equation eq.~\eqref{langevin quad} (or~\eqref{langevin quad annx}) 
can be obtained in the Fourier domain by means of the transfer function $S(\omega)$
\begin{align}
\hat{\mathbf{R}}_{\mathrm{out}}(\omega)=S(\omega)\hat{\mathbf{R}}_{\mathrm{in}}(\omega)
\end{align}
after using the in-out relation $\hat{\mathbf{R}}_{\mathrm{out}}(t)=\sqrt{2\Gamma}\hat{\mathbf{R}}(t)-\hat{\mathbf{R}}_{\mathrm{in}}(t)$,
where $S(\omega)$ is the matrix-valued function
\begin{equation}
S(\omega)=\sqrt{2\Gamma}\left(\mathrm{i}\omega\mathbb{I}+\Gamma-\mathcal{M}\right)^{-1}\sqrt{2\Gamma}-\mathbb{I}.
\label{S annx}
\end{equation}
Since $\mathcal{M}$ is Hamiltonian and $\Gamma$ is skew-Hamiltonian, we can prove~\cite{Gouzien2020} that $S(\omega)$ is $\omega$-symplectic~\cite{omegasymp},
so that $\hat{\mathbf{R}}_{\mathrm{out}}(\omega)$ are the Fourier transform of \textit{bona fide} boson quadrature operators.

\subsection{Higher-order morphing supermodes}\label{higherorder}
In the main text we illustrated only the first morphing supermodes.
In this section we report the structure of the 2nd, 3rd and 4th morphing supermodes for the different configurations discussed in the main text. 
Note that real and part imaginary part of $\bm{U}_{m}(\omega)$ are respectively symmetric and anti-symmetric with respect to $\omega$ as expected due to $\bm{R}_{\mathrm{out}}(\omega)$ symmetry properties. 
\begin{figure}[h!]
\includegraphics[width=\columnwidth]{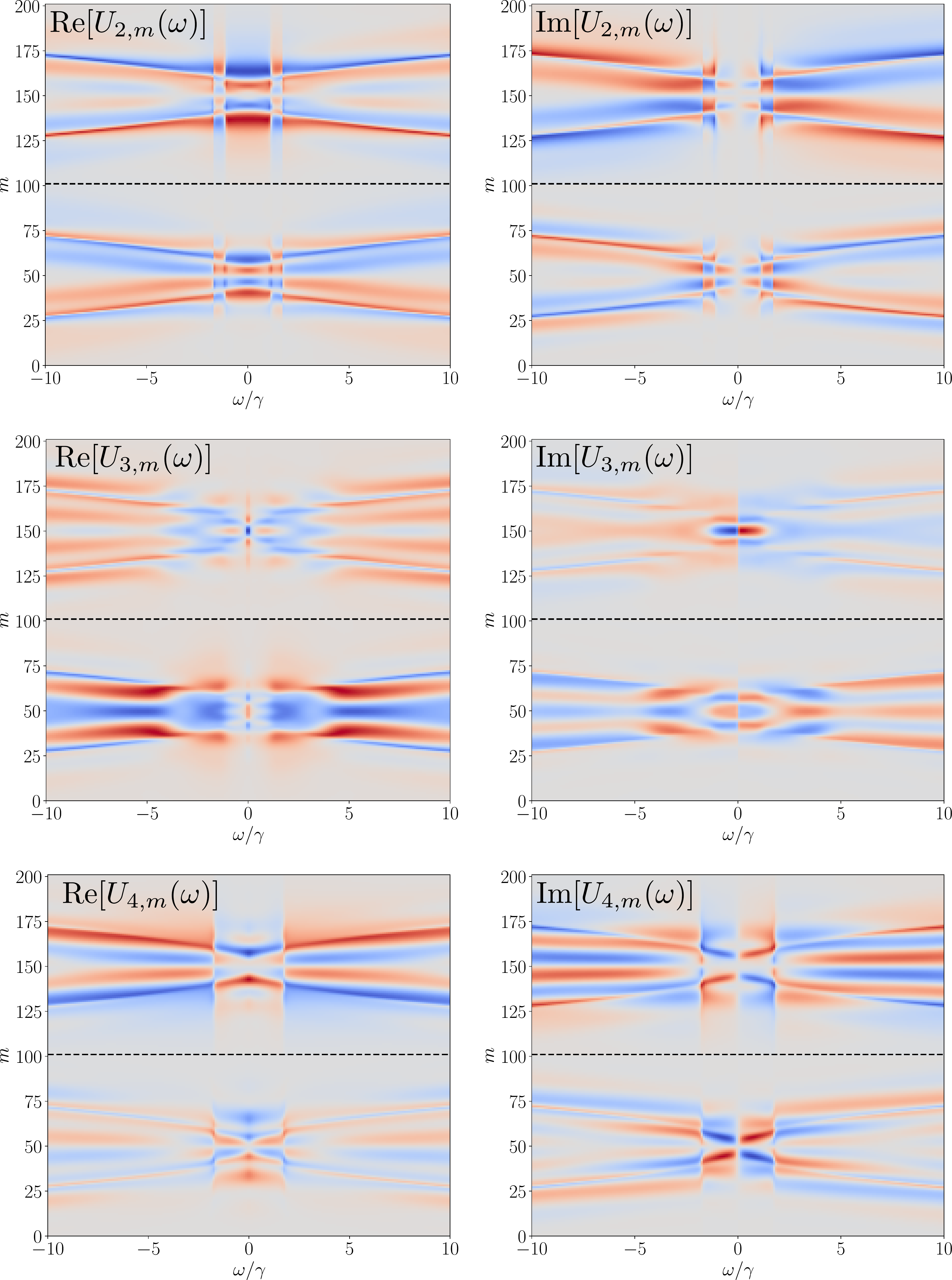}
\caption{Real and imaginary part of the second, third and fourth morphing supermodes corresponding to the configuration of Fig.~\ref{Figure3}.}\label{Fig:ms7}
\end{figure}
\begin{figure}[h!]
\includegraphics[width=\columnwidth]{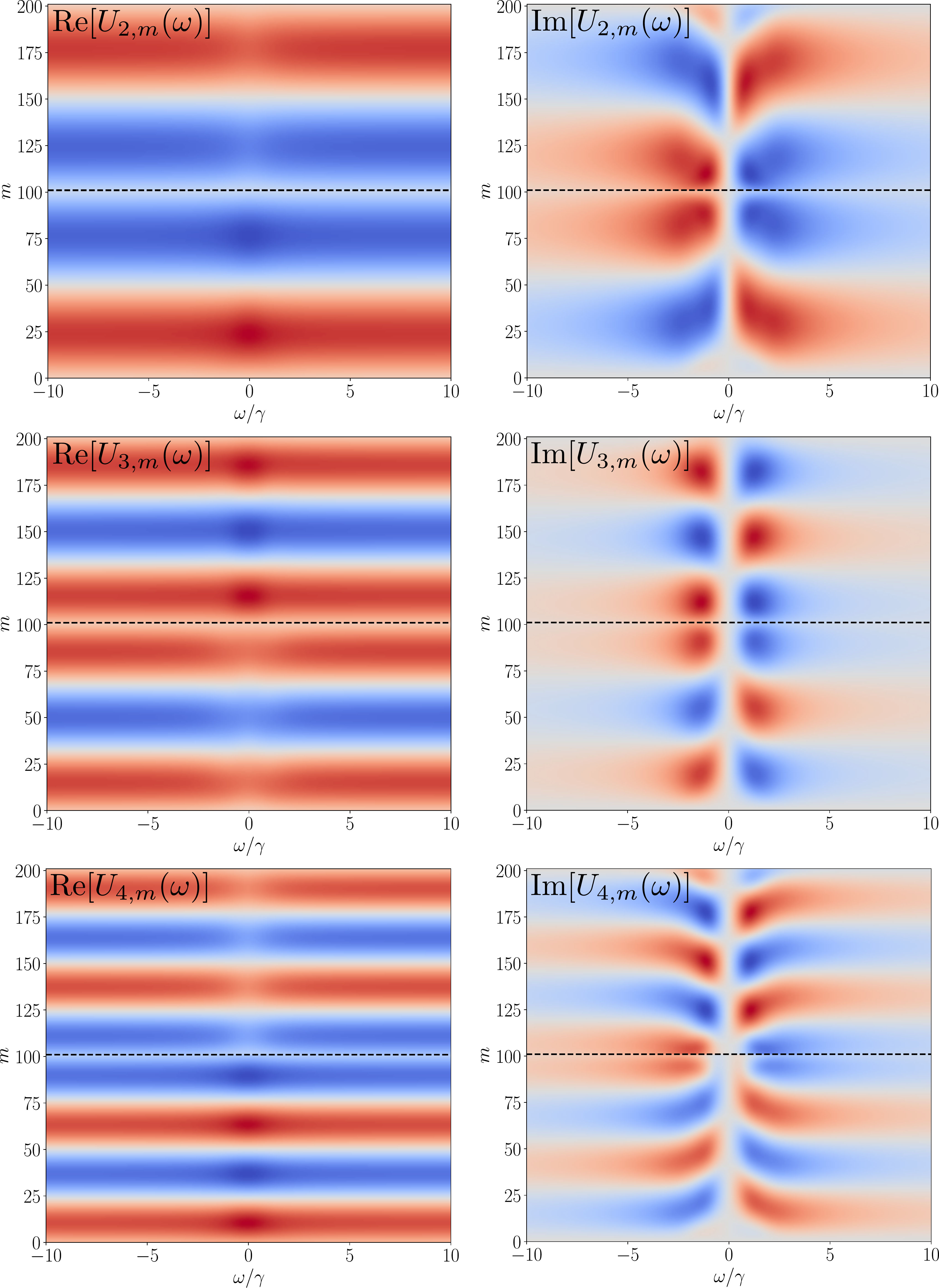}
\caption{Real and imaginary part of the second, third and fourth morphing supermodes corresponding to the configuration of Fig.~\ref{Figure:engineered}-top, without dispersion.}\label{Fig:E4}
\end{figure}
\begin{figure}[h!]
\includegraphics[width=\columnwidth]{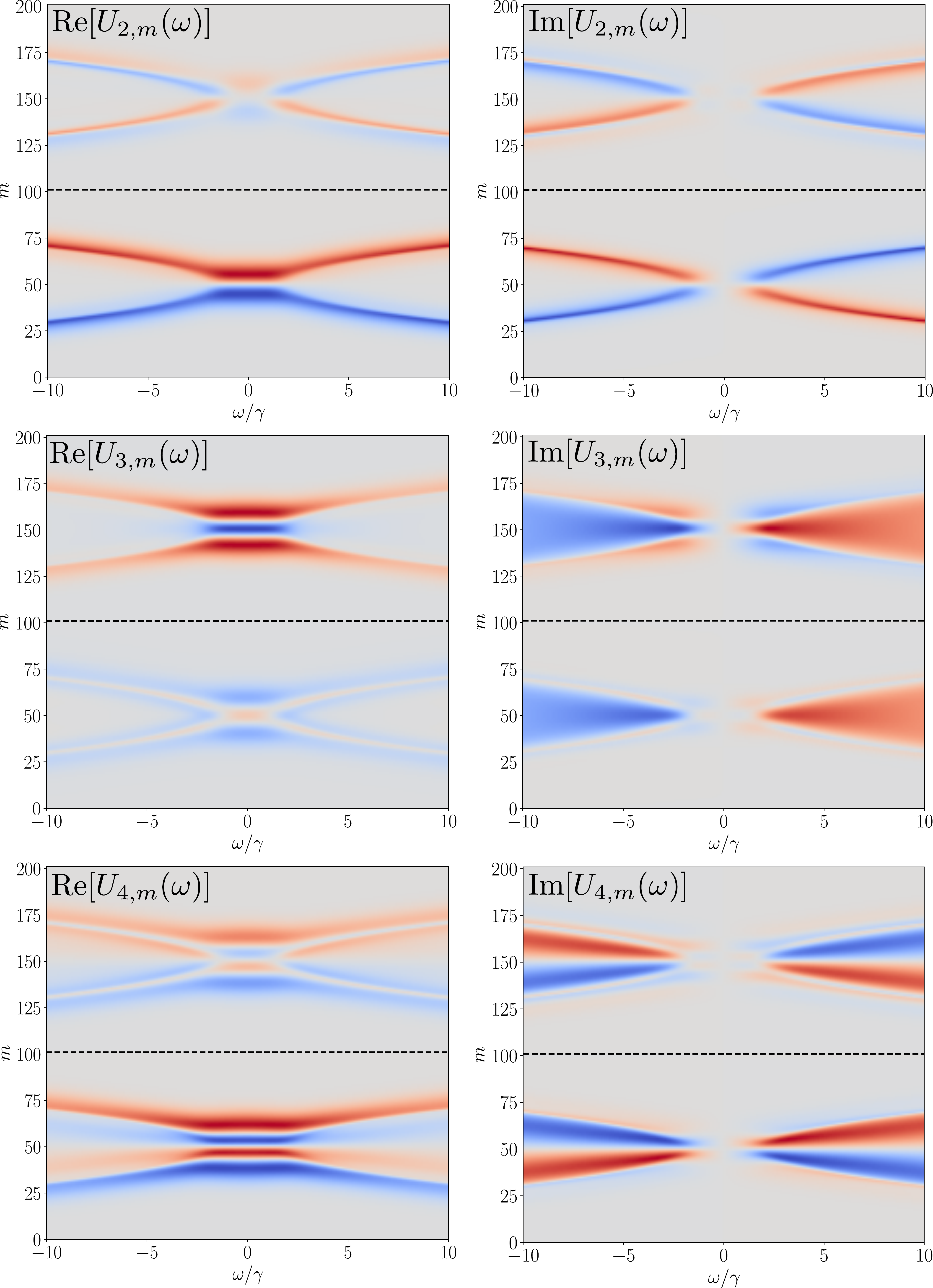}
\caption{Real and imaginary part of the second, third and fourth morphing supermodes corresponding to the configuration of Fig.~\ref{Figure:engineered}-bottom, with dispersion.}\label{Fig:E4bis}
\end{figure}
\FloatBarrier
\bibliography{biblio_ringSPOPO}

\begin{thebibliography}{32}
\bibitem [{qua()}]{quadfourier}%
  \BibitemOpen
  \href@noop {} {}\bibinfo {note} {In time domain the quadrature operators are
  defined such that their commutator is
  $[\hat{R}_{m}(t),\hat{R}_{n}(t')]=\mathrm{i}\Omega_{m,n}\delta(t-t')$, with
  $\Omega_{m,n}$ the elements of the symplectic form $\Omega$
  (see~\cite{omegasymp}). In Fourier domain
  $[\hat{R}_{m}(\omega),\hat{R}_{n}(\omega')]=\mathrm{i}\Omega_{m,n}\delta(\omega+\omega')$.}\BibitemShut
  {Stop}%
\end{thebibliography}

\end{document}